\documentclass[%
 amsmath,amssymb,
groupedaddress,
notitlepage,
]{revtex4-1}

\usepackage{graphicx}
\usepackage{dcolumn}
\usepackage{bm}
\usepackage{subfigure}

\begin{document}


\title{Detection of current induced spin polarization in epitaxial Bi$_2$Te$_3$ thin film}

\author{Rik Dey}
\email[]{rikdey@utexas.edu}

\author{Anupam Roy}
\author{Tanmoy Pramanik}
\author{Amritesh Rai}
\author{Seung Heon Shin}
\author{Sarmita Majumder}
\author{Leonard F. Register}
\author{Sanjay K. Banerjee}

\affiliation{Microelectronics Research Center, University of Texas at Austin, Austin, TX 78758, USA}

\date{\today}

\begin{abstract}
We electrically detect charge current induced spin polarization on the surface of molecular beam epitaxy grown Bi$_2$Te$_3$ thin film in a two-terminal device with a ferromagnetic MgO/Fe and a nonmagnetic Ti/Au contact. The two-point resistance, measured in an applied magnetic field, shows a hysteresis tracking the magnetization of the Fe. A theoretical estimate is obtained for the change in resistance on reversing the magnetization direction of Fe from coupled spin-charge transport equations based on quantum kinetic theory. The order of magnitude and the sign of the hysteresis is consistent with spin-polarized surface state of Bi$_2$Te$_3$.
\end{abstract}

\pacs{}

\maketitle

\nopagebreak

\par The three dimensional (3D) topological insulators (TIs) having insulating bulk and Dirac-type two dimensional (2D) surface states (SSs) with spin-momentum locking have potential for spintronics\cite{37,38,39,32,33,34}. The dispersion relation of the SS guarantees that any charge current flow within these states will induce a non-zero spin accumulation on the 2D surface of a 3D TI. This current induced spin polarization of the SS, controllable by the magnitude and the direction of the current, can be used to torque a ferromagnet (FM)\cite{32,33}. In recent experiments\cite{1,2,3,4,5,6,7,8,9,10,11,12,22}, the spin accumulation on the surface of 3D TIs Bi$_2$Se$_3$, (Bi$_x$Sb$_{1-x}$)$_2$Te$_3$, Bi$_{1.5}$Sb$_{0.5}$Te$_{1.7}$Se$_{1.3}$, BiSbTeSe$_{2}$, Bi$_2$Te$_2$Se and Sb$_2$Te$_3$, mostly grown by molecular beam epitaxy (MBE) or exfoliated, were electrically measured by the voltage probed with FM contact, where the voltage depends on the projection of SS spin polarization onto the FM magnetization direction. 

\par In this work, we detect the current induced spin polarization on the surface of an MBE grown Bi$_2$Te$_3$ thin film using Fe contact deposited on the surface and separated by a thin MgO barrier. We also provide a theoretical estimate of the detected spin signal, i.e., the voltage probed with the FM contact. Previously, the voltage drop measured between a FM and a nonmagnetic (NM) contact placed on the surface of a TI was theoretically calculated either using non-equilibrium Green's function\cite{13} or by solving the transport equations derived from Kubo formalism\cite{14}. Here, we provide a different approach for the derivation of the coupled spin-charge transport differential equation based on quantum kinetic theory\cite{15,16} in the diffusive limit. The experimentally measured spin signal matches well with the theory providing evidence for the spin polarized SS in our TI Bi$_2$Te$_3$ thin film.

\par The SSs of TIs are characterized by spin-momentum helically locked constant energy Fermi contour\cite{37,38,39}. However, due to band-bending near the surface, a 2D electron gas (2DEG) can be formed from the quantum confinement of the bulk states in the band-bending potential with Rashba spin splitting arising from the gradient of the confinement potential\cite{22,40,35,36}, and cannot be neglected \textit{a priori}. Therefore, we obtain coupled spin and charge transport equations for SSs of a TI as well as a Rashba 2DEG. For ease of analysis, we begin with the Rashba 2DEG, which is then modified for the TI SSs.
\par Within the parabolic band approximation, the Hamiltonian for the Rashba 2DEG (setting $\hbar = 1$) is\cite{15,17}:
\begin{equation}
\bm{H_{R}} = \frac{\vec{k}^2}{2m} \bm{\sigma_0}+ \lambda (\vec{{k}} \times \hat{{z}}) \cdot \vec{\bm{\sigma}}.
\end{equation}
Here $m$ is the effective mass, $\vec{k}$ is the in-plane momentum, $\bm{\sigma_0}$ is the identity matrix, $\vec{\bm{\sigma}} = (\bm{\sigma_x} \hat{x} + \bm{\sigma_y} \hat{y} + \bm{\sigma_z} \hat{z})$  where $\bm{\sigma}$'s are the Pauli spin matrices (we have used boldface for matrices in the spin space) and $\lambda$ is the strength of spin splitting. The spin-charge dynamic equation obtained from quantum kinetic theory can be written in terms of density matrix $\bm{\rho} = \rho_0 \bm{\sigma_0}+\vec{\rho} \cdot \vec{\bm{\sigma}}$ (where $\vec{\rho} = \rho_x \hat{x} + \rho_y \hat{y} + \rho_z \hat{z}$) as $\rho_r=\mathbb{D}_{rs}\rho_s$($r,s=0,x,y,z$), where $\mathbb{D}$ is the diffusion matrix\cite{15,16}. The charge and spin densities are given by $n=\rho_0$ and $\vec{s} =\vec{\rho}/2$, respectively. Considering uniform charge and spin densities along the $y$ direction, i.e., $\partial_y \bm{\rho} = 0$, the $4 \times 4$ diffusion matrix $\mathbb{D}$ is given by\cite{15,16}:
\begin{equation}
 \mathbb{D}=\int \frac{d\theta}{2\pi}
    \frac{\left[\begin{array}{cccc}
    \Omega(\Omega^2+\Omega_{SO}^2) & -i \sin\theta \cos\theta \Delta_x \Omega_{SO}^2 & i \Delta_x (\Omega^2+\Omega_{SO}^2 \cos^2\theta) & -i \sin\theta \Delta_x \Omega \Omega_{SO}	\\
   - i \sin\theta \cos\theta \Delta_x \Omega_{SO}^2 & \Omega(\Omega^2+\Omega_{SO}^2 \sin^2\theta+\Delta_x^2) & -\sin\theta \cos\theta \Omega \Omega_{SO}^2 & -\cos\theta (\Omega^2+\Delta_x^2)\Omega_{SO}	\\
    i \Delta_x (\Omega^2+\Omega_{SO}^2 \cos^2\theta) & -\sin\theta \cos\theta \Omega \Omega_{SO}^2 & \Omega(\Omega^2+\Omega_{SO}^2 \cos^2\theta) & -\sin\theta \Omega^2 \Omega_{SO}	\\
    i \sin\theta \Delta_x \Omega \Omega_{SO} & \cos\theta (\Omega^2+\Delta_x^2)\Omega_{SO} & \sin\theta \Omega^2 \Omega_{SO} & \Omega(\Omega^2+\Delta_x^2)
    \end{array}\right]}
    {\Omega^2(\Omega^2+\Omega_{SO}^2)+\Delta_x^2(\Omega^2+\Omega_{SO}^2 \cos^2\theta)}.
\end{equation}
with $\Omega = 1 - iw\tau + i v_m q_x  \tau \cos\theta$, $\Omega_{SO} = 2 k_F \lambda \tau$ and $\Delta_x = q_x \lambda \tau$. Here, $v_m = k_F/m$ is the Fermi velocity, $k_F$ is the Fermi momentum magnitude, $\tau$ is the momentum scattering time, $\theta$ is the angle between $\vec{k}$ and the $x$-direction, $w$ is the frequency of the temporal variation in the Fourier space and $q_x$ is the $x$-directional wave vector of the spatial variation in the Fourier space. The diffusion of the components of spin that are decoupled from the charge transport has been discussed in details previously\cite{15}. Here we are interested in the  coupled spin-charge transport in the diffusive limit, i.e., $w\tau \ll1$, $q_x l \ll 1$ and $k_F l \gg 1$, where $l=v_m \tau$ is the mean free path for the Rashba 2DEG. The Rashba spin splitting, which is due to the electric field from the gradient of the band-bending potential, will be significantly small as shown in literature\cite{22,40,*[{For Rashba 2DEG formed on the surface of a Bi$_2$Se$_3$ film with band bending of 0.13 eV over 20 nm distance\cite{40}, the splitting $\Delta k$ was calculated to be $0.004 \AA^{-1}$ which corresponds to $\lambda = (\hbar \Delta k/2m^*) = 1.7 \times 10^4$ m/s, assuming $m^*$ in Bi$_2$Se$_3$ is 0.13 times the free electron mass\cite{35}. To compare, the value of $v_F = 5 \times 10^5$ m/s in Bi$_2$Se$_3$ and $v_F = 4 \times 10^5$ m/s in Bi$_2$Te$_3$\cite{37,38,39}.}] [] 25}. So, we assume that the spin-splitting is much less than the Fermi energy, i.e., $\lambda \ll v_m$. Under these conditions, the terms $\mathbb{D}_{0x}, \mathbb{D}_{0z}, \mathbb{D}_{x0}, \mathbb{D}_{xy}, \mathbb{D}_{yx}, \mathbb{D}_{yz}, \mathbb{D}_{z0}, \mathbb{D}_{zy}$ are all zero. So, the transport of charge $n$ and $y$-component of spin $s_y$ are decoupled from the spin in the $x$ and $z$ directions when $\partial_y \bm{\rho} = 0$. To obtain the coupled dynamics of $n$ and $s_y$, we evaluate $\mathbb{D}_{00} \approx (1+iw\tau-q_x^2 l^2/2),\mathbb{D}_{0y} = \mathbb{D}_{y0} \approx -i q_x 2 \lambda \tau (\lambda k_F \tau)^2 , \mathbb{D}_{yy} \approx (1+iw\tau-q_x^2 l^2/2-2 (\lambda k_F \tau)^2)$. From $\rho_r=\mathbb{D}_{rs}\rho_s$($r,s=0,y$), the coupled diffusion equation for $n$ and $s_y$ becomes (with charge diffusion coefficient $D=v_m^2\tau/2$,  spin diffusion coefficient $D_s = D$, spin relaxation time $\tau_s = 2\tau/(2\lambda k_F \tau)^2$ and spin-charge coupling strength $\Gamma = -2\lambda (\lambda k_F \tau)^2$):
\begin{equation}
\begin{split}
& \partial_t n = D \partial_x^2 n + 2 \Gamma \partial_x s_y \\
& \partial_t s_y = D_s \partial_x^2 s_y -\frac{s_y}{\tau_s} + \frac{\Gamma}{2} \partial_x n.
\end{split}
\end{equation}

\par In case of SS of a TI, the Hamiltonian is\cite{14,16}:
 \begin{equation}
 \bm{H_{T}} = v_F (\vec{{k}} \times \hat{{z}}) \cdot \vec{\bm{\sigma}},
\end{equation}
with $v_F$ being the Fermi velocity of the SS. The Hamiltonian $\bm{H_T}$ can be obtained from $\bm{H_R}$ by the substitution $1/m \to 0$ and $\lambda = v_F$, so the corresponding matrix $\mathbb{D}$ will be given by Equation (2) with $\Omega = 1 - iw\tau $, $\Omega_{SO} = 2 k_F v_F \tau$ and $\Delta_x = q_x v_F \tau$. Similarly, in case of TI SSs\cite{16}, for $\partial_y \bm{\rho} = 0$ the spin dynamics in the $x$ and $z$ directions are decoupled from $n$ and $s_y$, while transport of n and $s_y$ are coupled. To get the spin-charge coupled diffusion equations, we evaluate the terms $\mathbb{D}_{00}, \mathbb{D}_{0y}, \mathbb{D}_{y0}, \mathbb{D}_{yy}$ under diffusive approximation, i.e. $w\tau \ll1$, $q_x l \ll 1$ and $k_F l \gg 1$, where $l=v_F \tau$ is the mean free path for the TI SS. Under these conditions, we obtain $\mathbb{D}_{00} \approx (1+iw\tau-q_x^2 l^2/2),\mathbb{D}_{0y} = \mathbb{D}_{y0} \approx i q_x l/2, \mathbb{D}_{yy} \approx \frac{1}{2}(1+iw\tau-3q_x^2 l^2/4)$ to the lowest order in $w \tau$ and $q_x l$. Therefore, the diffusion equation for $n$ and $s_y$ reads (with $D=v_F^2 \tau/2$, $D_s = 3D/2$, $\tau_s = \tau$ and $\Gamma = v_F/2$)\cite{23,*[{The spin relaxation time of the TI SS is typically in the order of 0.01-0.1 ps\cite{23}, same as the momentum scattering timescale\cite{28}}] [] 24}:
\begin{equation}
\begin{split}
& \partial_t n = D \partial_x^2 n + 2 \Gamma \partial_x s_y \\
& \partial_t s_y = D_s \partial_x^2 s_y -\frac{s_y}{\tau_s} + \Gamma \partial_x n.
\end{split}
\end{equation}
The spin-charge coupled transport for the Rashba 2DEG (Eq. 3) as well as for the SS of TI (Eq. 5) agree with that obtained previously from the Kubo formalism\cite{14,17}.

\begin{figure}
\includegraphics[scale=1]{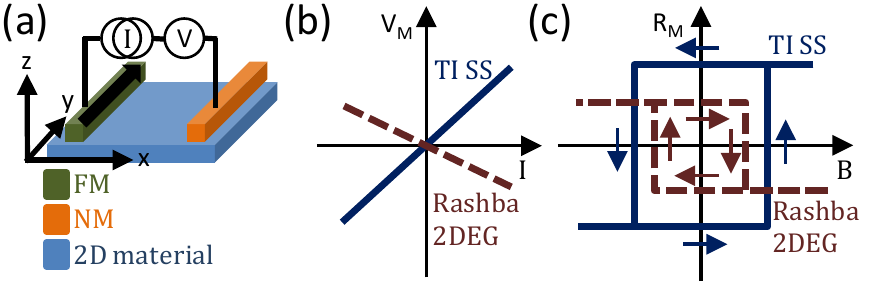}
\caption{\label{Hall_MR}(a) Schematic of a spin injection/detection experiment on the Rashba 2DEG or the TI SS, the current is passed through FM - NM contacts and voltage between them is measured. The detected voltage is $V=V_O+V_M$, where $V_O$ is the usual ohmic voltage drop and $V_M$ depends on the projection of the spin polarization onto the FM magnetization direction. (b) The variation of $V_M$ with injected current $I$ for FM magnetization along $+y$ direction. The slope $R_M$ is negative for the Rashba 2DEG and positive for TI SS. (c) The step-like hysteresis of $R_M$ with magnetic field (B) sweep that tracks the magnetization of the FM contact. The hysteresis is different for the Rashba 2DEG and the TI SS.}
\end{figure}

\par The signature of spin-charge coupled dynamics manifest itself as a magnetoresistance effect in a spin injection/detection experiment shown schematically in Fig. 1(a). In the shown experimental setup, one FM and one NM electrodes are deposited on the surface of the material along the $y$ direction, and a spin polarized current is injected into the 2D material through the left FM electrode in the $x$ direction. The FM is magnetized along the $y$ direction, and the voltage drop is measured between the two electrodes as the FM magnetization direction is reversed. In this measurement geometry, there will be no variation of the charge and the spin density along the $y$ direction, i.e., parallel to the electrodes, so the voltage drop can be calculated by solving the coupled diffusion equations, Eq. 3 or Eq. 5, with the proper boundary conditions for the charge and the spin currents. From the conservation of electric charge, the current continuity equation can be written as $\partial_t n+ \partial_x j_x = 0$, where $j_x$ is the particle current density ($1/e$ times the charge current density, where $e$ is the electric charge) along the $x$ direction, which is given by $j_x(x)=-D\partial_x n-2 \Gamma s_y$. So, for an injected current $I$, the boundary condition for the particle current density is $j_x(x=0)=j_x(x=L)=\frac{I}{eW}$, where $W$ is the width of the electrodes along the $y$ direction, and $L$ is the length between the two electrodes (assuming FM and NM electrodes are at $x=0$ and $x=L$ respectively). However, for the boundary condition of the $y$ component of the spin current density $j_s(x)$, only the contribution from the gradient of spin density should be considered, $j_s(x)=-D_s \partial_x s_y$\cite{14,17,18}. The FM injects a spin current density of $\frac{\eta m_y I}{eW}$ to the left, where $\eta$ is the density of state spin polarization of the FM and $m_y$ is the $y$-component of magnetization of the FM. The spin current extracted by the NM is zero for $L$ larger than the spin diffusion length, so we have $j_s(x=0)=\frac{\eta m_y I}{eW}$ and $j_s(x=L)=0$. In the static limit, the solution of the coupled diffusion equation with these boundary conditions will give the full electrochemical potential $n$, from which the voltage drop can be calculated as $V= - \frac{1}{e N_F} \int_0^L dx\text{ } \partial_x n$, where $N_F$ is the density of states at the Fermi level.

\par The voltage drop between the two electrodes consists of two parts, $V = V_O + V_M$, where the ohmic voltage drop $V_O \propto \frac{IL}{W}$ is independent of the magnetization of the FM, and the magnetoresistive part $V_M $ depends on the FM magnetization direction. For Rashba 2DEG, $V_M$ is approximately given by:
\begin{equation}
V_M = - \eta m_y R_\Box \frac{2\lambda \tau}{W} I,
\end{equation}
where $R_\Box = 1/\sigma$ is the sheet resistance, $\sigma = e^2 N_F D$ is the 2D conductivity. For the SS of TI, $V_M$ is given by:
\begin{equation}
V_M = + \eta m_y R_\Box \frac{v_F \tau}{2W} I.
\end{equation}
The voltage-current (V-I) characteristic is shown in Fig. 1(b) for FM magnetization along the $+y$ direction (i.e. $m_y = +1$) in case of both the Rashba 2DEG and the TI SSs. The voltage drop $V_M$ is linear with $I$, and the slope which is the resistance $R_M = V_M/I$ is negative for a Rashba 2DEG and positive for SSs of a TI. The opposite sign of the slope for the two cases is related to the sign of the spin-charge coupling strength $\Gamma$; $\Gamma$ is negative for the Rashba 2DEG and positive for the TI SS. As, typically\cite{25} $v_F$ is larger than $\lambda$, the slope $R_M$ is larger for TI SS than for Rashba 2DEG formed on the surface. The resistance between the FM and the NM electrodes will show a step-like hysteresis as the magnetic field direction is swept, as schematically shown in Fig. 1(b). For the Rashba 2DEG, a low resistance state will be observed for the $+$ve magnetic field and a high resistance state will be observed for the $-$ve magnetic field. However, for the TI SS, a higher resistance state will be detected for $+$ve field and a lower resistance state will be detected for $-$ve fields. The hysteresis of the resistance will follow the magnetization hysteresis of the FM electrodes (as $R_M \propto m_y$) with a coercive field value of the FM. The difference in the resistance $\Delta R = R_M(m_y=+1) - R_M(m_y=-1)$ is the measured spin signal in the experiment due to the current induced spin polarization of the TI SS.

\begin{figure}
\includegraphics[scale=1]{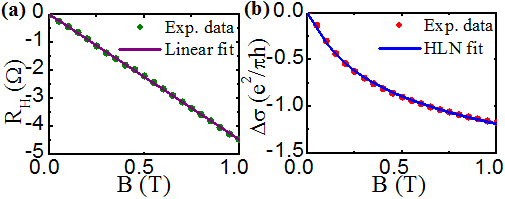}
\caption{\label{Hall_MR}(a) Hall resistance $R_H$ with magnetic field $B$ perpendicular to the sample surface at $2$ K, solid line is the linear fit. (b) Change in conductivity $\Delta \sigma$ with magnetic field $B$ perpendicular to the surface at $2$ K, solid line is the HLN fit.}
\end{figure}

\par In the experiment, we measure the spin signal for an MBE grown Bi$_2$Te$_3$ TI thin film of 4 nm thickness. The details of the growth and initial characterizations have been reported\cite{27,28}. The transport measurements are performed in a Physical Property Measurement System capable of cooling down to $2$ K with magnetic field up to $9$ T. The thin film is patterned in a Hall bar geometry with standard etching and lithography. The transverse Hall resistance and the longitudinal resistance are measured using  Ti/Au contacts deposited on the patterned thin film. Figure 2(a) shows the Hall resistance $R_H$ of the thin film at $2$ K with magnetic field perpendicular to the sample surface. The Hall resistance is linear with a negative slope, indicating the carriers are electrons. The 2D carrier concentration obtained from the slope of the linear fit is $n_{2D}=1.4 \times 10^{14}$ cm$^{-2}$. Such a high electron concentration indicates that both the surface and the bulk states are occupied\cite{27,28,29}. The conductivity $\sigma(B)$ obtained from the longitudinal resistance shows the signature of weak antilocalization (WAL). We plot in Fig. 2(b) the change in conductivity $\Delta \sigma(B) = \sigma(B) - \sigma(B=0)$ with magnetic field $B$ applied perpendicular to the surface at $2$ K. The sharp cusp near zero field is due to destruction of phase coherence of the electrons in an applied perpendicular field. The magnetoconductivity is explained with the Hikami-Larkin-Nagaoka (HLN) formula\cite{30}:
\begin{equation}
\Delta \sigma (B) = \alpha \frac{e^2}{\pi h} \left[ \psi \left( \frac{1}{2} + \frac{\hbar}{4el_\phi^2 B} \right) - \ln \left( \frac{\hbar}{4el_\phi^2 B} \right) \right],
\end{equation}
where $l_\phi$ is the phase coherence length, $h$ is the Planck's constant and $\alpha$ is a fitting parameter. From the HLN fitting shown in Fig. 2(b), we obtain $l_\phi = 121$ nm and $\alpha = - 0.46$. Although, the value of $\alpha \approx -1/2$ implies that both the surface and the bulk states are coupled and behave like a single phase coherent channel\cite{27,28}; it is possible that only the spin polarized surface state, and not the bulk states, contribute significantly to the spin signal\cite{2,5,6,7,22}. In our thin film, the 3D electron concentration $n_{3D} = 3.5 \times 10^{20}$ cm$^{-3}$ is close to the saturated electron concentration $n_{sat} = 4 \times 10^{20}$ cm$^{-3}$ in Bi$_2$Te$_3$ that corresponds to the stabilized Fermi level\cite{35}. As the bulk Fermi level is very close to the Fermi level stabilized on the surface, the band-bending near the surface will be small causing negligible Rashba spin splitting of the quantum confined bulk states that will not contribute to the spin signal. It was shown\cite{22} that, tuning Fermi level in the bulk gap will induce large band-bending near the surface that will cause large spin-splitting and give rise to opposite sign of the spin signal than that of TI SS.

\par To detect the spin signal in our thin film, we have fabricated a measurement geometry, shown in Fig. 1(a), of dimensions $L = 30 \text{ }\mu$m and $W = 35 \text{ }\mu$m with Fe as the FM and Ti/Au as the NM contact. We evaporate a patterned MgO(1 nm)/Fe(20 nm) stack on the top surface of the Bi$_2$Te$_3$ thin film. The thin layer of MgO helps in resolving the issue of resistance mismatch between the metallic Fe and the TI thin film, as well as protects the SS of the TI from the ferromagnetic exchange interaction that can break the time reversal symmetry. The Fe contact is rectangular in shape with the easy axis lying in the $y$-direction, and is capped with $21$ nm of Au. The magnetic field is applied parallel to the surface along the length of the Fe bar (along the easy axis), perpendicular to the direction of the current as shown in Fig. 1(a). Two terminal V-I measurements are recorded at each applied magnetic field as we sweep the field. The resistance $R$ at each magnetic field is obtained from the linear V-I characteristic, two of such data are shown in the insets of Fig. 3(a) and 3(b). Two sets of measurements are performed at $2$ K to obtain the resistance at different magnetic fields, one with the applied current ramped from zero to a positive value, and another with the current ramped from negative to a positive value. 

\begin{figure}
\includegraphics[scale=1]{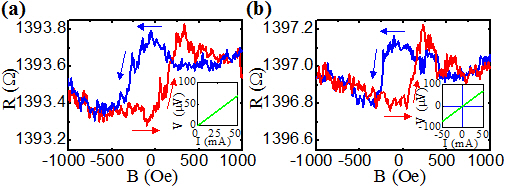}
\caption{\label{Hall_MR} Magnetic field ($B$) dependence of resistance $R$ between a FM and a NM contact deposited on the surface of the Bi$_2$Te$_3$ thin film, R obtained from $V-I$ characteristic at each field (one such measurement at zero field is shown in the inset of each figure) for current values of (a) $0 \text{ } \mu$A to $50 \text{ } \mu$A, (b) $-50 \text{ } \mu$A to $50 \text{ } \mu$A. The resistance shows hysteresis that mimics that of magnetization of the FM contact.}
\end{figure} 

\par Figure 3(a) shows the resistance $R$ with the magnetic field sweep from a positive field of $1000$ Oe to a negative field of $-1000$ Oe and back to a positive field of $1000$ Oe. In the first set of measurements shown in Fig. 3(a), the resistance is obtained from the V-I characteristic with current values of $0 \text{ } \mu$A to $50 \text{ }\mu$A. The inset of Fig. 3(a) shows one such V-I plot at zero field. From the hysteresis of resistance with applied field shown in Fig. 3(a), it is seen that a high resistive state is obtained at positive magnetic field, while a low resistive state is obtained at negative field. This is consistent with the theory which predicts the same resistive state for the TI SSs, but a different resistive state for the Rashba 2DEG, as shown in Fig. 1(c). Similar hysteresis is observed in the second set of measurements as shown in Fig. 3(b), where we have obtained the resistance from V-I characteristic with current values of $-50 \text{ } \mu$A to $50 \text{ }\mu$A. The inset of Fig. 3(b) shows one such V-I plot at zero magnetic field. As seen in Fig. 3(b), the resistance is higher for positive magnetic fields and lower for negative fields, consistent with that of Fig. 3(a). The hysteresis resembles the one in Fig. 3(a) showing a similar coercive field value. The hysteresis loop observed in the resistance versus applied magnetic field is almost square shape with a coercive field value of about $250$ Oe. The hysteresis loop is shifted towards a negative field value, which can be due to the exchange bias between ferromagnetic Fe and the anti-ferromagnetic oxide of Fe. Also, the local peaks seen in the resistance near the coercive field values can be attributed to the magnetic domain reversal in the multi-domain Fe contact\cite{2}. However, the hysteresis overall shows a single step-like behavior when the FM magnetization direction changes from $+y$ to $-y$ and vice-versa. The observed hysteresis in the resistance measured both for positive and negative currents show that the magnetoresistive voltage drop is directly proportional to the magnitude and the sign of the applied current giving a linear V-I characteristics. This consideration rules out the possibility that the observed hysteresis is due to the anisotropic magnetoresistance or the anisotropic tunnel magnetoresistance of the Fe contacts, as the voltage drop due to these effects are only proportional to the magnitude of current and not the direction\cite{2,3,5}. Also, the Hall effect from the film due to the perpendicular component of fringe fields of the Fe, or the spin Hall effect of the bulk can be excluded as these effects will give rise to a signal much smaller than that we have observed\cite{2,3,5}. Further, the anomalous Hall effect or the anomalous Nernst effect of the Fe contact is excluded by doing a controlled experiment. The hysteresis of resistance with magnetic field matches with magnitude (that we show next) and the sign of that for the TI SS while being opposite in sign and higher in magnitude than that of Rashba 2DEG. So, we exclude the possibility that the spin signal is due to a Rashba 2DEG formed at the surface or the interface. Hence, both sets of measurement, shown in Fig. 3, indicate that the observed spin signal is indeed of the nature predicted in theory due to the SS of the Bi$_2$Te$_3$ thin film. 

\par We obtain the value of $\Delta R = 0.3 \text{ } \Omega$ from the experiment. The theoretical value of $\Delta R$ for TI SSs can be calculated from Eq. 7, using $\sigma = \frac{e^2}{2h} k_F l$ for the TI SSs:
\begin{equation}
\Delta R = \eta \frac{2h}{e^2} \frac{1}{k_F W}.
\end{equation}
From the carrier concentration\cite{*[{Assuming two 2D surface states and at least two degenerate quasi-2D (because of our thin film) bulk states being populated, i.e. at least 4 states being filled, an upper limit of $k_F$ is estimated from the carrier concentration $n_c$ using $k_F=\sqrt{\pi n_c}$}] [] 26}, the estimated $k_F = 2$ nm$^{-1}$. Using the values $W=35\text{ }\mu$m and $\eta \approx 0.45$ for Fe\cite{2,10}, we obtain $\Delta R \approx 0.33\text{ }\Omega$. The theoretical estimate agrees well with the experimental value, indicating that SS in the thin film  mainly contribute to the spin signal. 

\par In conclusion, we derived the coupled spin-charge transport equations from quantum kinetic theory for both the Rashba 2DEG and the TI SS. Solving the differential equations with proper boundary conditions, the behavior of the resistance between a FM and a NM contact is obtained as a function of the magnetization of the FM. We experimentally measured the resistance between a FM and a NM contact on the surface of a MBE grown Bi$_2$Te$_3$ thin film, which shows hysteresis with the applied field tracking that of the magnetization of the FM contact. The experimental value of the difference in resistance on reversing the FM magnetization direction agrees well in magnitude and sign with that of a TI SS, providing evidence of spin polarized SS in the thin film.


\par This work is supported by the NRI SWAN and the NSF NNCI program.

\newpage

\section*{Supplementary Material}

\section*{S1: Theoretical model}\label{Sec1}

\par Here, we are going to derive the coupled spin and charge transport equations for the Rashba two-dimensional electron gas (2DEG) as well as the surface state (SS) of a topological insulator (TI) from quantum kinetic equation under the diffusive approximation. We consider a general Hamiltonian for 2DEG with Rashba spin splitting ($\hbar = 1$)\cite{15,17}:
\begin{equation}
\tag{S1}
 \bm{H_{R}} = \frac{\vec{k}^2}{2m} \bm{\sigma_0} + \lambda (\vec{{k}} \times \hat{{z}}) \cdot \vec{\bm{\sigma}},
\end{equation}
where $m$ is the effective mass, $\vec{k} = (k_x \hat{x} + k_y \hat{y}) = k \hat{k}$ is the in-plane momentum ($k=\lvert\vec{k}\rvert$), $\bm{\sigma_0}$ is the identity matrix, $\vec{\bm{\sigma}} = (\bm{\sigma_x} \hat{x} + \bm{\sigma_y} \hat{y} + \bm{\sigma_z} \hat{z})$  with $\bm{\sigma}$'s being the Pauli matrices (bold letter indicating a matrix in spin space) and $\lambda$ is the strength of the SOC.  The Hamiltonian for the SS of a TI reads\cite{14,16}:
 \begin{equation}
\tag{S2}
 \bm{H_{T}} = v_F (\vec{{k}} \times \hat{{z}}) \cdot \vec{\bm{\sigma}},
\end{equation}
with $v_F$ being the Fermi velocity of the SS. As $\bm{H_T}$ can be obtained from $\bm{H_R}$ by taking $1/m \to 0$ and $\lambda = v_F$, we will write down the kinetic equations for the more general Hamiltonian $\bm{H_R}$ first and then obtain the one for $\bm{H_T}$ afterwards by the appropriate substitution.

\par The quantum kinetic equation can be written in terms of angular distribution function $\bm{g}(\theta,\vec{R},T)$ (where $\theta = \tan^{-1}(k_y/k_x)$, $\vec{R}$ is the in-plane position and $T$ is the time) and density matrix $\bm{\rho}(\vec{R},T) = \int \frac{d\theta}{2\pi} \bm{g}(\theta,\vec{R},T)$ as\cite{15,16}:
\begin{equation}
\tag{S3}
 \partial_T \bm{g} + \frac{1}{2} \{\overline{\vec{\bm{V}}} \cdot \vec{\nabla}_{\vec{R}}, \bm{g} \} + i \left[ \overline{\bm{H}}, \bm{g} \right] + \frac{\bm{g}}{\tau} = \frac{\bm{\rho}}{\tau}.
\end{equation}
Here, $\bm{H}$ is any Hamiltonian, $\vec{\bm{V}} = \partial \bm{H} / \partial \vec{k}$ is the velocity operator, and $\tau$ is the momentum scattering time assuming random spin-independent delta-correlated impurity potential. The quantities $\overline{\vec{\bm{V}}}$ and $\overline{\bm{H}}$ are averages over the energy and are functions of $(\theta,\overline{k})$, where the average of any function $F(\vec{k})$ is $\overline{F}(\theta,\overline{k}) = \int - d\epsilon_k f^{\prime}_{FD}(\epsilon_k)F(\vec{k})$ with $f_{FD}(\epsilon_k)$ being the Fermi-Dirac distribution and $\epsilon_k$ being the energy dispersion of the Hamiltonian $\bm{H}$. At zero temperature, $\overline{F}$ peaks the value at the constant energy Fermi surface, so becomes only function of $\theta$ and the Fermi momentum $k_F$ (as $\overline{k} \to k_F$, the Fermi wave vector, at zero temperature). By writing $\bm{g} = g_0 \bm{\sigma_0}+\vec{g} \cdot \vec{\bm{\sigma}}$  (where $\vec{g} = (g_x \hat{x} + g_y \hat{y} + g_z \hat{z})$) and $\bm{\rho} = \rho_0 \bm{\sigma_0}+\vec{\rho} \cdot \vec{\bm{\sigma}}$ (where $\vec{\rho} = (\rho_x \hat{x} + \rho_y \hat{y} + \rho_z \hat{z})$), and taking trace of Equation (S3) after multiplying by $\sigma_i$, where $i=0,x,y,z$, and using the fact that $Tr(\sigma_i\sigma_j)=2\delta_{ij}$, we will get:
\begin{equation}
\tag{S4}
\begin{split}
\Bigg[\bigg( \partial_T+\frac{1}{\tau}+\frac{\overline{k}}{m} \hat{k} \cdot \vec{\nabla}_{\vec{R}} \bigg)g_0 + \lambda \partial_y g_x - \lambda \partial_x g_y \Bigg]=\frac{\rho_0}{\tau},\\
\Bigg[\bigg( \partial_T+\frac{1}{\tau}+\frac{\overline{k}}{m} \hat{k}  \cdot \vec{\nabla}_{\vec{R}} \bigg)g_x + \lambda \partial_y g_0 + 2 \lambda \overline{k} (\hat{k} \cdot \hat{x}) g_z \Bigg]=\frac{\rho_x}{\tau},\\
\Bigg[\bigg( \partial_T+\frac{1}{\tau}+\frac{\overline{k}}{m} \hat{k} \cdot \vec{\nabla}_{\vec{R}} \bigg)g_y - \lambda \partial_x g_0 + 2 \lambda \overline{k} (\hat{k} \cdot \hat{y}) g_z \Bigg]=\frac{\rho_y}{\tau},\\
\Bigg[\bigg( \partial_T+\frac{1}{\tau}+\frac{\overline{k}}{m} \hat{k} \cdot \vec{\nabla}_{\vec{R}} \bigg)g_z - 2 \lambda \overline{k} (\hat{k} \cdot \hat{x}) g_z - 2 \lambda \overline{k} (\hat{k} \cdot \hat{y}) g_z \Bigg]=\frac{\rho_z}{\tau}.
\end{split}
\end{equation}

\par Equation (S4) can be written in a matrix form in the Fourier space $(\vec{q},w)$ of $(\vec{R},T)$ (Fourier transformed $\partial_T \to -iw$ and $\vec{\nabla}_{\vec{R}} \to i\vec{q}$)  as $\mathbb{K}_{rs}g_s=\rho_r$ ($r,s=0,x,y,z$) with the $4\times4$ matrix $\mathbb{K}$ given by: 

\begin{equation}
\tag{S5}
\mathbb{K} =   \left[\begin{array}{cccc}
 \Omega & i\Delta_y & -i\Delta_x & 0	\\
  i\Delta_y & \Omega & 0 & \Omega_{SO}\cos\theta \\
 -i\Delta_x & 0 & \Omega &  \Omega_{SO}\sin\theta	\\
 0 & -\Omega_{SO}\cos\theta & -\Omega_{SO}\sin\theta & \Omega
    \end{array}\right].
\end{equation}

\par Here, $\Omega = 1 - iw\tau + i (\frac{\overline{k}}{m}) (\hat{k} \cdot \vec{q})$, $\Omega_{SO} = 2 \lambda \overline{k} \tau$ and $\Delta_{x,y} = \lambda q_{x,y} \tau$. There were typos in previous reports\cite{15,16} in the literature that we have corrected here in Equation (S4), (S5), and (S6) (Equation (2) in the main article). Now, we consider uniform charge and spin density along the $y$ direction, i.e. $\partial_y \bm{\rho}=0 \implies q_y=0$, so $\mathbb{K}^{-1}$ becomes:

\begin{equation}
\tag{S6}
    \frac{\left[\begin{array}{cccc}
    \Omega(\Omega^2+\Omega_{SO}^2) & -i \sin\theta \cos\theta \Delta_x \Omega_{SO}^2 & i \Delta_x (\Omega^2+\Omega_{SO}^2 \cos^2\theta) & -i \sin\theta \Delta_x \Omega \Omega_{SO}	\\
   - i \sin\theta \cos\theta \Delta_x \Omega_{SO}^2 & \Omega(\Omega^2+\Omega_{SO}^2 \sin^2\theta+\Delta_x^2) & -\sin\theta \cos\theta \Omega \Omega_{SO}^2 & -\cos\theta (\Omega^2+\Delta_x^2)\Omega_{SO}	\\
    i \Delta_x (\Omega^2+\Omega_{SO}^2 \cos^2\theta) & -\sin\theta \cos\theta \Omega \Omega_{SO}^2 & \Omega(\Omega^2+\Omega_{SO}^2 \cos^2\theta) & -\sin\theta \Omega^2 \Omega_{SO}	\\
    i \sin\theta \Delta_x \Omega \Omega_{SO} & \cos\theta (\Omega^2+\Delta_x^2)\Omega_{SO} & \sin\theta \Omega^2 \Omega_{SO} & \Omega(\Omega^2+\Delta_x^2)
    \end{array}\right]}
    {\Omega^2(\Omega^2+\Omega_{SO}^2)+\Delta_x^2(\Omega^2+\Omega_{SO}^2 \cos^2\theta)}.
\end{equation}

\par Using the fact that $\bm{g}=\mathbb{K}^{-1}\bm{\rho}$  and $\bm{\rho} = \int \frac{d\theta}{2\pi} \bm{g}$, the spin-charge dynamic equation can be written as $\rho_r=\mathbb{D}_{rs}\rho_s$, where $\mathbb{D} = \int \frac{d\theta}{2\pi} \mathbb{K}^{-1}$. It was shown that\cite{15,16}, in case of $q_y=0$, the diffusion of the $x$- and $z$-components of spin are decoupled from the charge and $y$-component of spin transport. So, we are interested in the $\mathbb{D}_{00}, \mathbb{D}_{0y}, \mathbb{D}_{y0}, \mathbb{D}_{yy}$ components of the diffusion matrix that will give rise to spin-charge coupled diffusion equations.


\subsection{Rashba 2DEG}

\par Assuming $q_y=0$, we obtain $\Omega = 1 - iw\tau + i v_m q_x  \tau \cos\theta$, $\Omega_{SO} = 2 k_F \lambda \tau$ and $\Delta_x = q_x \lambda \tau$, where, $v_m = k_F/m$ is the Fermi velocity. The terms $\mathbb{D}_{0x}, \mathbb{D}_{0z}, \mathbb{D}_{x0}, \mathbb{D}_{xy}, \mathbb{D}_{yx}, \mathbb{D}_{yz}, \mathbb{D}_{z0}, \mathbb{D}_{zy}$ are all zero in the diffusive limit, i.e., under the conditions $w\tau \ll1$, $q_x l \ll 1$, $k_F l \gg 1$ ($l=v_m \tau$ is the mean free path for the Rashba 2DEG), and with the assumption that the spin splitting is smaller than the Fermi energy, i.e., $\lambda \ll v_M$. These conditions also implies $\Delta_x \ll q_x l \ll 1$ and $\Delta_x \ll \lvert \Omega \rvert$ (as $\Omega = (1-iw\tau+iq_xl \cos\theta)\approx1$ since $w\tau \ll1$, $q_x l \ll 1$). With these assumptions, we show that:

\begin{equation}
\tag{S7}
\begin{split}
\mathbb{D}_{0x} = \mathbb{D}_{x0} & = \int \frac{d\theta}{2\pi} \frac{-i \sin\theta \cos\theta \Delta_x \Omega_{SO}^2}{\Omega^2(\Omega^2+\Omega_{SO}^2)+\Delta_x^2(\Omega^2+\Omega_{SO}^2 \cos^2\theta)} \\
& \approx \int \frac{d\theta}{2\pi} \frac{-i \sin\theta \cos\theta \Delta_x \Omega_{SO}^2}{\Omega^2(\Omega^2+\Omega_{SO}^2)} \Bigg( 1 - \frac{\Delta_x^2(\Omega^2+\Omega_{SO}^2 \cos^2\theta)}{\Omega^2(\Omega^2+\Omega_{SO}^2)} \Bigg) \\
& \approx \int \frac{d\theta}{2\pi} \frac{-i \sin\theta \cos\theta \Delta_x \Omega_{SO}^2}{\Omega^2(\Omega^2+\Omega_{SO}^2)} \\
& \approx \int \frac{d\theta}{2\pi} (-i) \sin\theta \cos\theta \Delta_x  \Bigg[\frac{1}{\Omega^2}-\frac{1}{\Omega^2+\Omega_{SO}^2}\Bigg] \\
& \approx \int \frac{d\theta}{2\pi} (-i) \sin\theta \cos\theta \Delta_x  \Bigg[\frac{1}{1-2iw\tau+2iq_x l \cos\theta}-\frac{1}{1-2iw\tau+2iq_x l \cos\theta+\Omega_{SO}^2}\Bigg] \\
& \approx 0 \text{ (upto first order in $w$ and $q_x$)}.
\end{split}
\end{equation}
Similarly, $\mathbb{D}_{xy}= \mathbb{D}_{yx}, \mathbb{D}_{0z}= - \mathbb{D}_{z0}, \mathbb{D}_{yz}= - \mathbb{D}_{zy}$ are all zero in the diffusive approximation. So we have obtained that the $x$- and $z$-components of spins are decoupled from charge and $y$-component of spin. Now, to get the spin-charge coupled transport, we evaluate the other terms:

\begin{equation}
\tag{S8}
\begin{split}
\mathbb{D}_{00}& = \int \frac{d\theta}{2\pi} \frac{\Omega(\Omega^2+\Omega_{SO}^2)}{\Omega^2(\Omega^2+\Omega_{SO}^2)+\Delta_x^2(\Omega^2+\Omega_{SO}^2 \cos^2\theta)} \\
& \approx \int \frac{d\theta}{2\pi} \frac{\Omega(\Omega^2+\Omega_{SO}^2)}{\Omega^2(\Omega^2+\Omega_{SO}^2)} \Bigg( 1 - \frac{\Delta_x^2(\Omega^2+\Omega_{SO}^2 \cos^2\theta)}{\Omega^2(\Omega^2+\Omega_{SO}^2)} \Bigg)  \\
& \approx \int \frac{d\theta}{2\pi} \frac{1}{(1-iw\tau+iq_xl\cos\theta)} \\
& \approx \int \frac{d\theta}{2\pi} (1+iw\tau-iq_xl\cos\theta) \\
& \approx \Big(1+iw\tau-\frac{q_x^2 l^2}{2}\Big).
\end{split}
\end{equation}
Similarly, $\mathbb{D}_{0y} = \mathbb{D}_{y0} \approx -i q_x 2 \lambda \tau (\lambda k_F \tau)^2 $ and $\mathbb{D}_{yy} \approx (1+iw\tau-q_x^2 l^2/2-2 (\lambda k_F \tau)^2)$. 

\par From $\rho_r=\mathbb{D}_{rs}\rho_s$($r,s=0,y$), the spin-charge coupled diffusion equation can be written as $(\bm{I}-\bm{D})\bm{\rho_2}=0$, where $\bm{I}$ is a 2$\times$2 identity matrix, $\bm{D}$ is a 2$\times$2 diffusion matrix for charge and spin. The charge density $n=\rho_0$ and spin density $\vec{s} = \vec{\rho}/2$, so $\bm{\rho_2}=(n\text{ }2s_y)^T$. The coupled diffusion equation in Fourier space becomes:
\begin{equation}
\tag{S9}
\left[\begin{array}{cc}
- iw\tau+\frac{q_x^2 l^2}{2} & iq_x 2 \lambda \tau (\lambda k_F \tau)^2\\
iq_x 2 \lambda \tau (\lambda k_F \tau)^2 & -iw\tau+q_x^2 l^2/2+2 (\lambda k_F \tau)^2
    \end{array}\right] 
\Bigg[\begin{array}{c}
n\\
2s_y
    \end{array}\Bigg] = 0.
\end{equation}
\par Now, inverse Fourier transforming back to time and real-space variation (i.e. $-iw \to \partial_t$ and $i\vec{q_x} \to \partial_x$), we get:
\begin{equation}
\tag{S10}
\begin{split}
& \partial_t n = D \partial_x^2 n + 2 \Gamma \partial_x s_y \\
& \partial_t s_y = D_s \partial_x^2 s_y -\frac{s_y}{\tau_s} + \frac{\Gamma}{2} \partial_x n,
\end{split}
\end{equation}
whcih matches with the one derived from Kubo formalism\cite{17}. Here, charge diffusion coefficient $D=v_m^2\tau/2$,  spin diffusion coefficient $D_s = D$, spin relaxation time $\tau_s = 2\tau/(2\lambda k_F \tau)^2$ and spin-charge coupling strength $\Gamma = -2\lambda (\lambda k_F \tau)^2$.


\subsection{Surface state of a TI}

\par The Hamiltonian $\bm{H_T}$ of the SS of a TI, given by Equation (S2), can be obtained from Equation (S1) by the substitution $1/m \to 0$ and $\lambda = v_F$, so the corresponding matrix $\mathbb{K}^{-1}$ will be given by Equation (S6) with $\Omega = 1 - iw\tau $, $\Omega_{SO} = 2 v_F k_F \tau$ (as $\overline{k} \approx k_F$ at 2 K) and $\Delta_x = v_F q_x \tau$. As the denominator of $\mathbb{K}^{-1}$ is a function of $\cos^2\theta$, by the symmetry of the trigonometric function in the four quadrants ($\theta \in [0,\pi/2), [\pi/2, \pi), [\pi, 3\pi/2), [3\pi/2, 2\pi))$, the terms $\mathbb{D}_{0x}, \mathbb{D}_{0z}, \mathbb{D}_{x0}, \mathbb{D}_{xy}, \mathbb{D}_{yx}, \mathbb{D}_{yz}, \mathbb{D}_{z0}, \mathbb{D}_{zy}$ are all zero after angular integration. So, the spin dynamics in the $x,z$ directions are decoupled from charge transport, while the charge and $y$-component of spin are coupled. To get the spin-charge coupled dynamics, we evaluate the terms $\mathbb{D}_{00}, \mathbb{D}_{0y}, \mathbb{D}_{y0}, \mathbb{D}_{yy}$ under diffusive approximation, i.e. $w\tau \ll1$ and $q_x l \ll 1 \implies \Delta_x \ll 1$ ($l=v_F \tau$ is the mean free path for the SS, in this case), and $k_F l \gg 1$  i.e. $\Omega_{SO} \gg 1$ ($\implies \Omega_{SO} \gg \lvert \Omega \rvert, \Delta_x$). Under these conditions, we obtain:

\begin{equation}
\tag{S11}
\begin{split}
\mathbb{D}_{00}& = \int \frac{d\theta}{2\pi} \frac{\Omega(\Omega^2+\Omega_{SO}^2)}{\Omega^2(\Omega^2+\Omega_{SO}^2)+\Delta_x^2(\Omega^2+\Omega_{SO}^2 \cos^2\theta)} \\
& \approx \int \frac{d\theta}{2\pi} \frac{\Omega \Omega_{SO}^2}{\Omega^2\Omega_{SO}^2 + \Delta_x^2\Omega_{SO}^2 \cos^2\theta} \\
& \approx \int \frac{d\theta}{2\pi} \frac{1-iw\tau}{1-2iw\tau+\Delta_x^2 \cos^2\theta} \\
& \approx \int \frac{d\theta}{2\pi} (1-iw\tau)(1+2iw\tau-\Delta_x^2 \cos^2\theta) \\
& \approx \Big(1+iw\tau-\frac{\Delta_x^2}{2}\Big).
\end{split}
\end{equation}
Similarly,
\begin{equation}
\tag{S12}
\begin{split}
\mathbb{D}_{0y} = \mathbb{D}_{y0} & = \int \frac{d\theta}{2\pi} \frac{i\Delta_x(\Omega^2+\Omega_{SO}^2\cos^2\theta)}{\Omega^2(\Omega^2+\Omega_{SO}^2)+\Delta_x^2(\Omega^2+\Omega_{SO}^2 \cos^2\theta)} \\
& \approx \int \frac{d\theta}{2\pi} \frac{i\Delta_x \Omega_{SO}^2\cos^2\theta}{\Omega^2\Omega_{SO}^2 + \Delta_x^2\Omega_{SO}^2 \cos^2\theta} \\
& \approx \int \frac{d\theta}{2\pi} i\Delta_x\cos^2\theta(1+2iw\tau-\Delta_x^2 \cos^2\theta) \approx \frac{i\Delta_x}{2},
\end{split}
\end{equation}
and
\begin{equation}
\tag{S13}
\begin{split}
\mathbb{D}_{yy}& = \int \frac{d\theta}{2\pi} \frac{\Omega(\Omega^2+\Omega_{SO}^2\cos^2\theta)}{\Omega^2(\Omega^2+\Omega_{SO}^2)+\Delta_x^2(\Omega^2+\Omega_{SO}^2 \cos^2\theta)} \\
& \approx \int \frac{d\theta}{2\pi} \frac{\Omega \Omega_{SO}^2\cos^2\theta}{\Omega^2\Omega_{SO}^2 + \Delta_x^2\Omega_{SO}^2 \cos^2\theta} \\
& \approx \int \frac{d\theta}{2\pi} \frac{(1-iw\tau)\cos^2\theta}{1-2iw\tau+\Delta_x^2 \cos^2\theta} \\
& \approx \int \frac{d\theta}{2\pi} (1-iw\tau)(1+2iw\tau-\Delta_x^2 \cos^2\theta)\cos^2\theta \\
& \approx \frac{1}{2}\Big(1+iw\tau-\frac{3\Delta_x^2}{4}\Big).
\end{split}
\end{equation}

So, the matrix diffusion equation in Fourier space reads:
\begin{equation}
\tag{S14}
\left[\begin{array}{cc}
- iw\tau+\frac{\Delta_x^2}{2} & -\frac{i\Delta_x}{2} \\
-\frac{i\Delta_x}{2} & \frac{1}{2}-\frac{iw\tau}{2}+\frac{3\Delta_x^2}{8}
    \end{array}\right] 
\Bigg[\begin{array}{c}
n\\
2s_y
    \end{array}\Bigg] = 0.
\end{equation}
\par Now, using the value of $\Delta_x=q_x l$, and inverse Fourier transforming back to time and real-space variation (i.e. $-iw \to \partial_t$ and $i\vec{q_x} \to \partial_x$), we get the coupled diffusion equation:
\begin{equation}
\tag{S15}
\begin{split}
& \partial_t n = D \partial_x^2 n + 2 \Gamma \partial_x s_y \\
& \partial_t s_y = D_s \partial_x^2 s_y -\frac{s_y}{\tau_s} + \Gamma \partial_x n,
\end{split}
\end{equation}
where, $D=v_F^2 \tau/2$, $D_s = 3D/2$, $\tau_s = \tau$ and $\Gamma = v_F/2$. These coupled equations are same as obtained previously\cite{14}.


\subsection{Solution of diffusion equations}

\par In the static limit, we write the coupled spin-charge diffusion equation for both the Rashba 2DEG and the TI SS in the following general form:
\begin{equation}
\tag{S16}
\begin{split}
&  D_c \partial_x^2 n + \Gamma_c \partial_x s_y = 0\\
&  D_s \partial_x^2 s_y -\frac{s_y}{\tau_s} + \Gamma_s \partial_x n = 0,
\end{split}
\end{equation}
with $D_i$ and $\Gamma_i$ ($i=c,s$) replaced by appropriate values for Rashba 2DEG or TI SS in specific cases.

\par Now, we solve the equation for a charge current $I$ injected along the $+x$-direction with a finite width $W$ of the injecter contacts. As, the particle current density along the $x$ direction is given by $j_x(x)=-D\partial_x n- \Gamma_c s_y$, and the injected particle current density is ($\frac{I}{eW}$), we get from the first of the Equation (S16):
\begin{equation}
\tag{S17}
\begin{split}
&  D_c \partial_x n + \Gamma_c s_y = -\frac{I}{eW}\\
\implies & \partial_x n = \frac{1}{D_c} \bigg(-\frac{I}{eW} - \Gamma_c s_y \bigg).
\end{split}
\end{equation}
Inserting Equation (S17) into the second of Equation S(16), we get:
\begin{equation}
\tag{S18}
\begin{split} 
& D_s \partial_x^2 s_y -\frac{s_y}{\tau_s} + \frac{\Gamma_s}{D_c} \bigg(-\frac{I}{eW} - \Gamma_c s_y \bigg) = 0 \\
\implies & D_s \partial_x^2 s_y -\frac{s_y}{\tau_s'} - \frac{\Gamma_s I}{e W D_c} = 0,
\end{split}
\end{equation}
where we have introduced $\frac{1}{\tau_s'} = \frac{1}{\tau_s} + \frac{\Gamma_s \Gamma_c}{D_c}$. Now, we follow the derivation on Ref.\cite{18} [5], and make the substitution $s_y^1 = s_y + \frac{\Gamma_s I \tau_s'}{e W D_c}$, to get the new differential equation $D_s \partial_x^2 s_y^1 -\frac{s_y^1}{\tau_s'} = 0$, or $s_y^1 =\tau_s' D_s \partial_x^2 s_y^1$. Using Equation (S17), we calculate the voltage drop between the two contacts situated at a distance $L$ as:
\begin{equation}
\tag{S19}
\begin{split} 
& V= - \frac{1}{e N_F} \int_0^L dx\text{ } \partial_x n  = \frac{1}{e N_F} \int_0^L dx\text{ } \frac{1}{D_c} \bigg(\frac{I}{eW} + \Gamma_c \Big( s_y^1 - \frac{\Gamma_s I \tau_s'}{e W D_c}\Big) \bigg)\\
\implies &V= \frac{IL/W}{e^2 N_F D_c} \Big(1- \frac{\Gamma_c \Gamma_s \tau_s'}{D_c}\Big) -\Gamma_c \tau_s' \int_0^L dx\text{ } (-D_s \partial_x^2 s_y^1) = V_O + V_M.
\end{split}
\end{equation}
The first term gives the ohmic voltage drop $V_O \propto (IL/W)$ and the second term is due to the spin current injection. In the boundary condition of the $y$ component of the spin current density $j_s(x)$ , only the contribution from the gradient of spin density should be considered, $j_s(x)=-D_s \partial_x s_y$\cite{14,17,18}. The FM injects a spin current density of $\frac{\eta m_y I}{eW}$ at $x=0$ and no spin current is injected or extracted by the NM at $x=L$ assuming $L$ (order of $\mu$m) is much larger than spin diffusion length (order of nm). So, using $j_s(x=0)=\frac{\eta m_y I}{eW}$ and $j_s(x=L)=0$ in Equation (S19), we get:
\begin{equation}
\tag{S20}
V_M = \frac{\eta m_y I/W}{e^2 N_F D_c} \Gamma_c \tau_s' =  \eta m_y R_\Box \frac{I}{W} \Gamma_c \tau_s' ,
\end{equation}where$R_\Box = 1/\sigma$ is the sheet resistance, $\sigma = e^2 N_F D_c$ is the 2D conductivity.

\par For, Rashba 2DEG, using $\Gamma_c = 2 \Gamma $ and $\Gamma_s = \Gamma /2$ (where $\Gamma = -2\lambda (\lambda k_F \tau)^2$)and $D_c=D=v_m^2 \tau/2$, we get $1/\tau_s' = 1/\tau_s + 8 m^2 \lambda^2 (\lambda k_F \tau)^2/k_f^2$ implying $\tau_s' \approx \tau_s$ (as $\lambda \ll v_m = k_F/m$). Now using the value of  $\tau_s = 2\tau/(2\lambda k_F \tau)^2$ and $\Gamma_c$, we obtain: 
\begin{equation}
\tag{S21}
V_M = - \eta m_y R_\Box \frac{2\lambda \tau}{W} I.
\end{equation}
\par For the TI SS, using $\Gamma_c = 2 \Gamma $, $\Gamma_s = \Gamma$ (where $\Gamma = v_F/2$), $D_c=D=v_F^2 \tau/2$ and $\tau_s = \tau$, we get $\tau_s' = \tau/2$ and
\begin{equation}
\tag{S22}
V_M = +\eta m_y R_\Box \frac{v_F \tau}{2W} I.
\end{equation}


\section*{S2: Controlled experiments}\label{Sec2}

\begin{figure}
\subfigure{\includegraphics[scale=0.3]{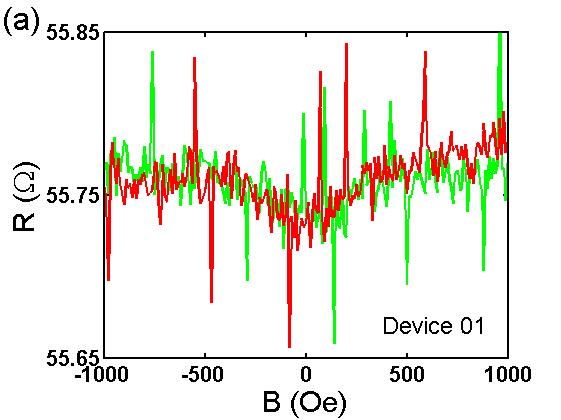}}
\subfigure{\includegraphics[scale=0.3]{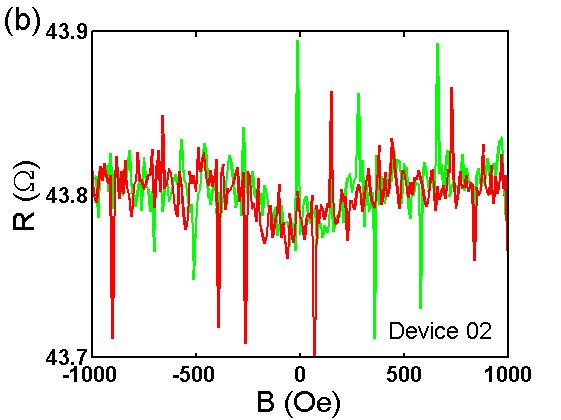}}
\renewcommand\thefigure{S1}
\caption{\label{figS1} Two-point magnetoresistance between a ferromagnetic MgO/Fe and a nonmagnetic Ti/Au contact deposited on the surface of a 10 nm thick Au film shows no hysteresis. Data for two separate devices are given in (a) and (b).}
\end{figure}

\par As a supporting evidence to show that the observed spin signal is indeed arising from the Bi$_2$Te$_3$ thin film, and not due to other spurious effects such as anisotropic magnetoresistance or anisotropic tunnel magnetoresistance or the anomalous Hall effect or the anomalous Nernst effect in the ferromagnetic Fe contact, we repeat the experiments on Au film of 10 nm thickness. We  fabricated identical two probe device with ferromagnetic MgO/Fe and nonmagnetic Ti/Au contacts of the same dimensions and thicknesses of the ones we used for measurement on Bi$_2$Te$_3$. As shown in Figure S1, the measured resistance in an applied magnetic field is parabolic, which is a characteristic of the Au film, and no hysteresis with field sweep is observed. This results confirms none of the spurious effects in Fe is responsible for the observed hysteresis in the resistance with magnetic field sweep in our Bi$_2$Te$_3$ sample. So, we conclude that the hysteresis observed in the measurement on the TI thin film is due to the spin polarized surface state of Bi$_2$Te$_3$.

\end{document}